# EVALUATING OPEN ACCESS PAPER REPOSITORY IN HIGHER EDUCATION FOR ASEAN REGION

Reza Chandra
Gunadarma University, Indonesia
reza_chan@staff.gunadarma.ac.id

Arif Purwo Nugroho
Gunadarma University, Indonesia
arifpurwo@staff.gunadarma.ac.id

Fikri Saleh
Gunadarma University, Indonesia
fikri@staff.gunadarma.ac.id

## ABSTRACT

Paper repository at higher education is a collection of scientific articles created by the academic society. This study took as many as 80 universities in the Webometrics ranking of repositories in the Southeast Asia region. The tools used in this research is Google for number of web page and Google Scholar for number of document paper repository and Ahrefs for referring page, backlink and reffering domain. The result of this study, Eprints is the most widely used tools in higher education, as many as 37 higher educations (46,25%). Institut Teknologi Sepuluh November got the highest score in number of web page in Google (2.010.000), Bogor Agricultural University Scientific Repository got the highest score for number of document paper (44.300). University of Sumatera Utara Repository got the highest score for reffering page (82588) and backlink (86421). Universiti Teknologi Malaysia Institutional Repository got the highest score for reffering domain (532).

---

## 1. INTRODUCTION

Paper repository at higher education is a collection of scientific articles created by the academic society. Paper repository contains journals, proceedings, thesis, dissertations and other scientific articles. Paper repository also serves as digital libraries needed by researchers in the search for scientific reference source.

To facilitate academic society in developing their knowledge, open access policies created for scientific work. Patel Research [1] suggests the presence of a policy of open access to higher education, research results can be publicized more widely, plagiarism can be reduced because the public can see it. Higher education also benefited from research results can be seen to add to the reputation and credibility of the college.

Open access paper repository policy in higher education make research produced by higher education become a reference so that the higher education website traffic increases. Many institutions linking the paper repository portal on their portal to facilitate their researchers for finding references.

The search engine also plays an important role in optimizing the portal repository paper. For example, the Google search engine and Google Scholar. Most of the researchers looking for references through a search engine. It is to impact the traffic of paper repository portal. The more visitors to the portal repository paper, the greater the traffic portals and research of higher education increasingly publicized so that the effect of plagiarism is reduced.

This study attempts to evaluating of the paper repository portal for the ASEAN region indexed in the Webometrics Ranking Web of Repositories January 2014 based on its tools and popularity on the web.

## 2. THEORETICAL BACKGROUND

Based on repositories.webometrics.info, The Ranking Web of World repositories is an initiative of the Cybermetrics Lab, a research group belonging to the Consejo Superior de Investigaciones Científicas (CSIC), the largest public research body in Spain. CSIC is attached to the Ministry of Education and its main objective is to promote scientific research as to improve the progress of the scientific and technological level of the country which will contribute to increase the welfare of the citizens.

The organization collaborates with other institutions of the Spanish R&D system (universities, autonomous governs, other public and private research organisms) and with social, economic, national or foreign agents to which contributes with its research capacity and human and material resources in the development of research projects or under the form of consultancy and scientific and technical support. CSIC was founded in 1939 from a previous body, the Junta para la Ampliación de Estudios e Investigaciones Científicas created in 1907 under the leadership of the Spanish Nobel Prize Prof. Ramón y Cajal.

The Cybermetrics Lab using quantitative methods has designed and applied indicators that allow us to measure the scientific activity on the Web. The cybermetric indicators are useful to evaluate science and technology and they are the perfect complement to the results obtained with bibliometric methods in scientometric studies.

According to Abrizah, Noorhidawati and Kiran [2], there are several studies on paper repository as in 10 European countries, namely Belgium, France, the United Kingdom (UK), Denmark, Norway, Sweden, Finland, Germany, Italy and the Netherlands but the research only focus on on acquisition of content almost exclusively on faculty publications.

Studies of Chen and Hsiang [3] on Asian institutional repositories revealed that Open Access repositories are not widespread and about 4-10 percent has a centralized institutional repositories for about 300 universities except in mainland China. The big contributors to the growth of institutional repositories in Asia are Japan, India and Taiwan.

Wani, Gul and Rah [4] [5] [6] [7] [8] [9]in October 2008 findings the most used software was DSpace (90 countries) followed by EPrints (15 countries). Journal articles in English were the most prominent content type deposited.

## 3. METHODS

The sample was 80 higher educations in South East Asia included in the ranking of repositories based on their activity on the webometrics. Measurements on the first stage is to check whether the higher education has its own repository website.

The second stage is to examine the software that used in the paper repository website, number of web page, number of document paper repository.

The third stage is to check paper repository website popularity. Paper repository website popularity is measured using Ahrefs for referring page, backlink and reffering domain. Observation and measurements conducted research variables in the same week that in early May 2014.

Description of the variables are presented graphically to determine the pattern of utilization of paper repository. The influence of ownership paper repository websites and various career website features tested with independent sample t test and regression analysis linking with referring domains and total backlinks paper repository with reffering web domain and total backlinks college website.

## 4. RESULT AND DISCUSSION

According to webometrics ranking web of repositories published on Januari 2014, Eprints is the most used software in South East Asia. There are 37 colleges using Eprints (46,25%) followed by DSpace, 27 colleges (33,75%), other software 14 colleges (17,5%), Open Journal System, 0 college (0%).

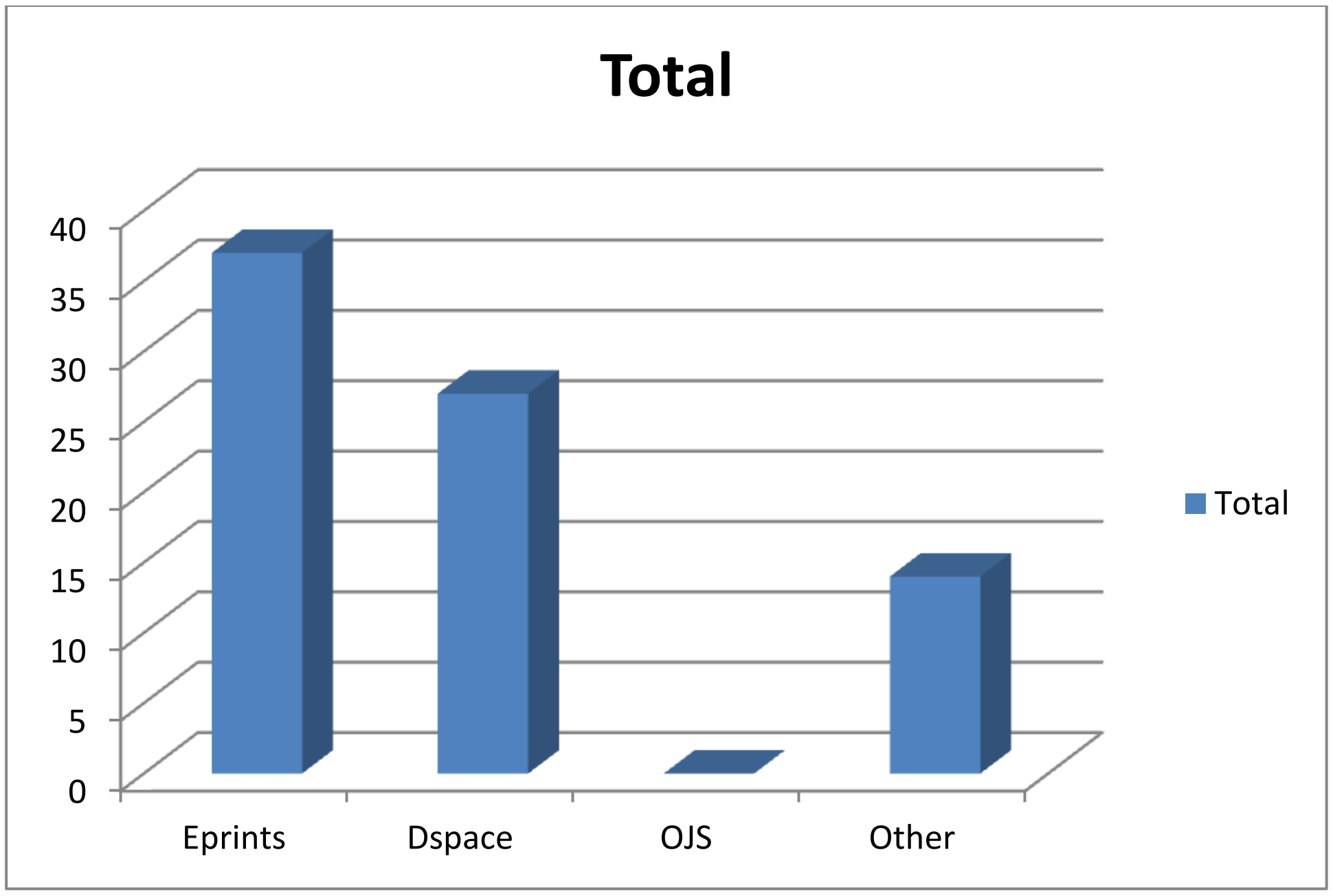

**Figure 1.** Paper Repository Software

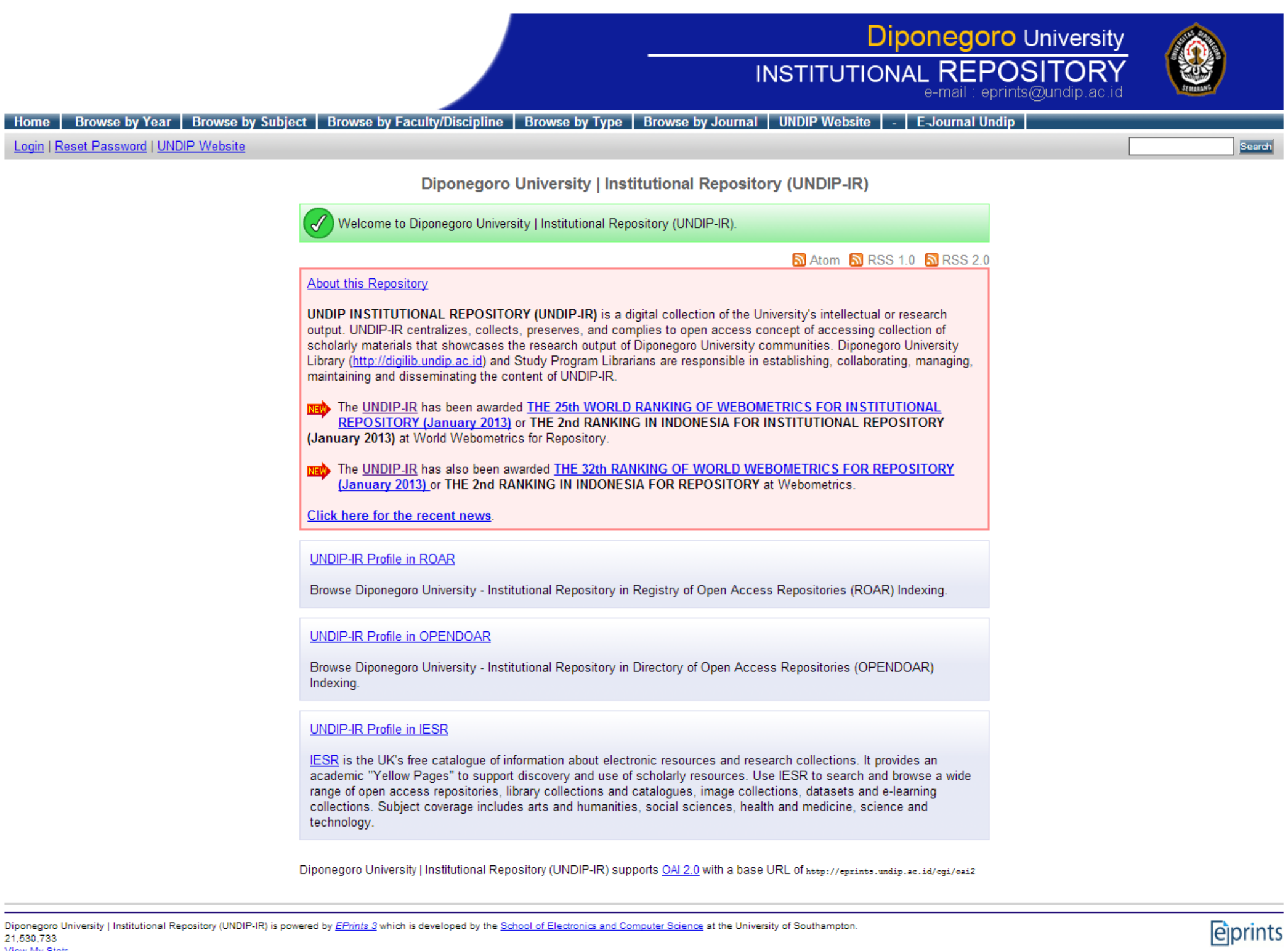

**Figure 2.** Paper Repository using EPrints

In 80 college indexed webometrics in January 2014, there are 2 websites that are offline, in example repository.upnyk.ac.id and repository.stisitelkom.ac.id

Indonesia is the big contributors to the growth of institutional repositories in South East Asia (43 colleges) followed by Malaysia (21 colleges), Thailand (9 colleges), Philipines (4 colleges), Singapore (2 colleges) and Vietnam (1 college).

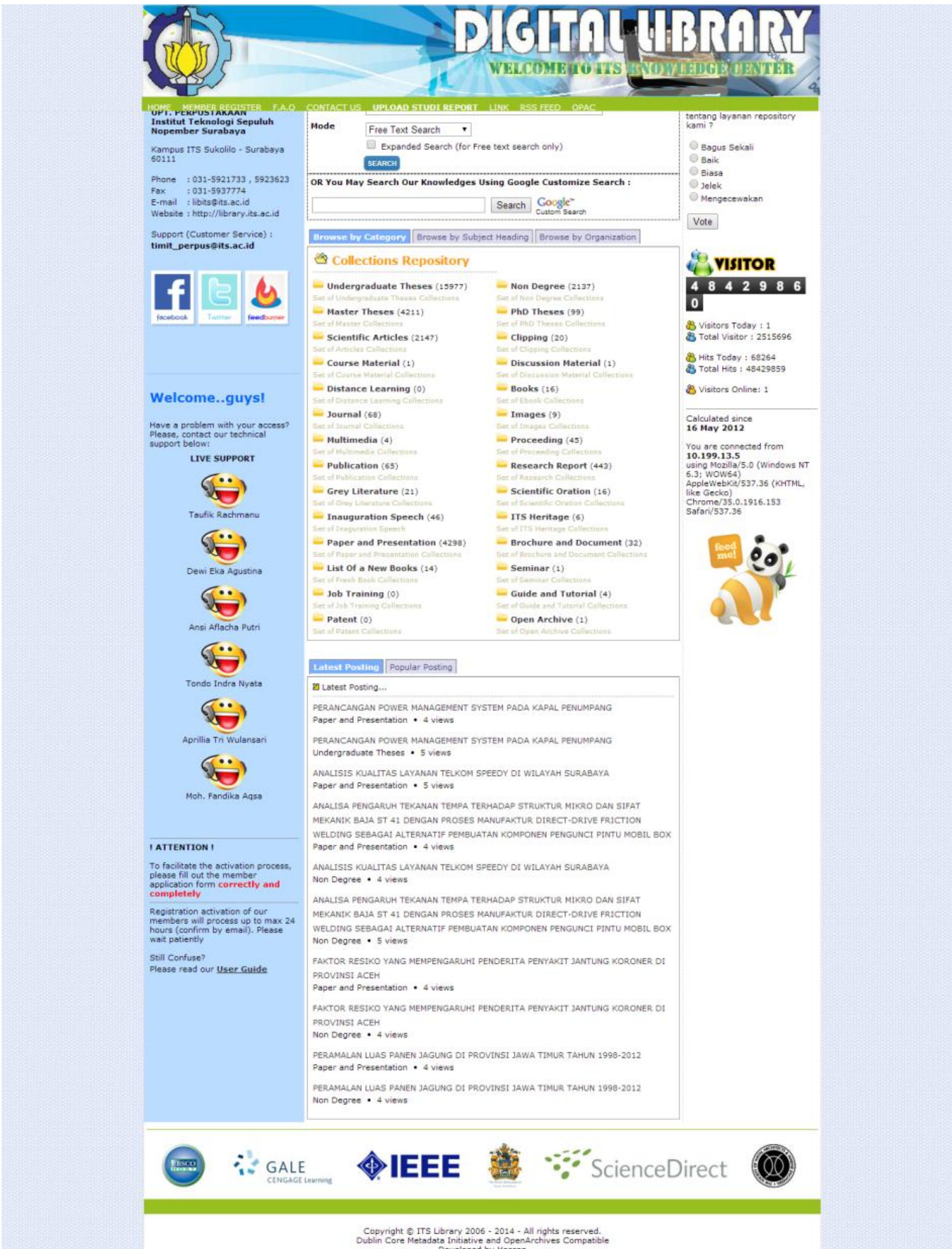

**Figure 3.** Institut Teknologi Sepuluh Nopember Repository from Indonesia Got Number 1 Position in Webometrics Rank of Repositories

The top 3 colleges in number of page Google is from Indonesia. They are, Institut Teknologi Sepuluh November (2.010.000), Universitas Muhammadiyah Surakarta (1370000) and Bogor Agricultural Institute (1350000). In 80 paper repositories website, PDF is the most uploaded file type in Google followed by Docx and Doc.

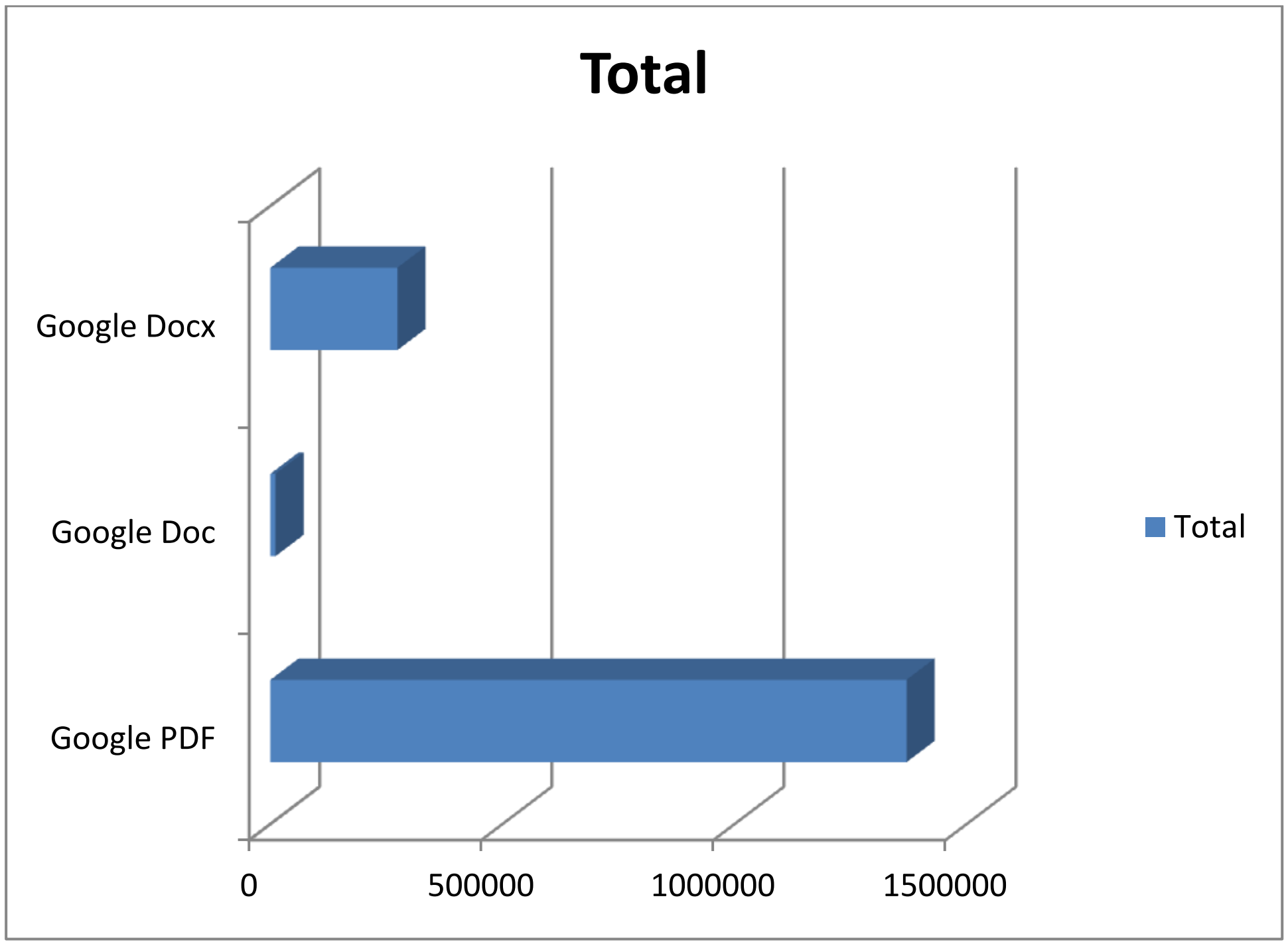

**Figure 4.** Rich File Type in Google

Meanwhile, in number of page Google Scholar the top 3 colleges are Bogor Agricultural Institute (44300), Diponegoro University (33000) and Universiti Teknologi Malaysia (23500).

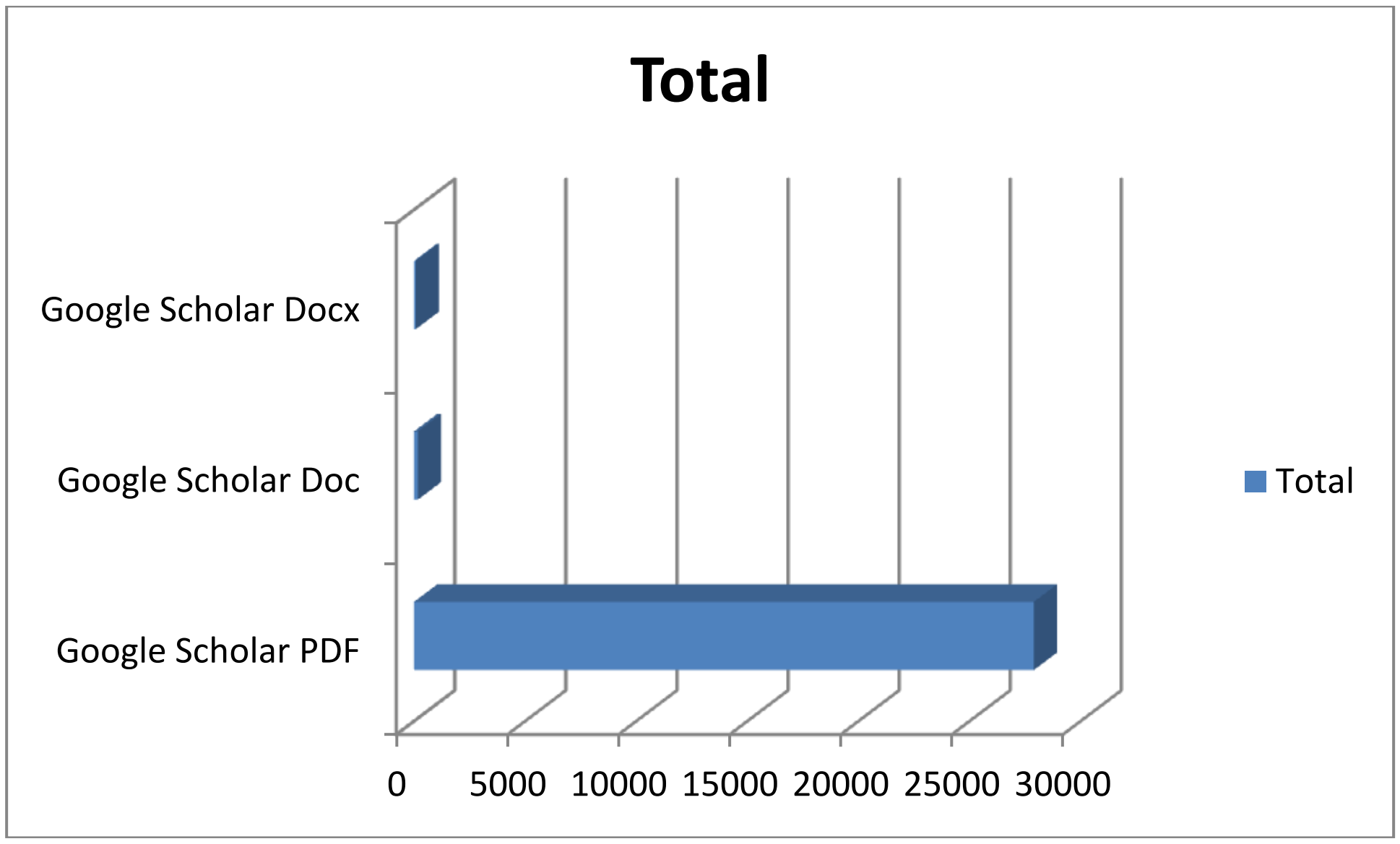

**Figure 5.** Rich File Type in Google Scholar

Measuring the impact of the paper repository consists of two parts, namely the impact of ownership on the paper repository website popularity college. Tests of significance using independent sample t test. The results of the test to the table below.

**Table 1.** Result Of T Test

| | Test Value = 0 | | | | | |
|---|---|---|---|---|---|---|
| | | | | | 95% Confidence Interval of the Difference | |
| | t | df | Sig. (2-tailed) | Mean Difference | Lower | Upper |
| Referring Web Page | 2.998 | 79 | .004 | 3264.82500 | 1097.3012 | 5432.3488 |
| Backlinks Web | 3.544 | 79 | .001 | 4438.72500 | 1945.5339 | 6931.9161 |
| Referring Domain Web | 7.667 | 79 | .000 | 92.63750 | 68.5864 | 116.6886 |

## 5. CONCLUSION

According to webometrics ranking web of repositories published in South East Asia edition January 2014, EPrints is the most used software in college, Indonesia got the most college in top 80 repositories in South East Asia. PDF is the most file type uploaded in paper repository. Ownership paper repository website got positive impact on college website popularity.